\documentclass[12pt]{article}
\usepackage[top = 2 cm, bottom = 2 cm, left = 2.3 cm, right = 2 cm]{geometry}
\usepackage{authblk, cite, color, amssymb, amsmath}
\usepackage[hypertex, colorlinks = true, linkcolor = blue, citecolor = red]{hyperref}
\usepackage[dvipsnames]{xcolor}

\begin{document}

\title{Probing the Information-Probabilistic Description}

\author[1,2]{Merab Gogberashvili}
\author[3]{Beka Modrekiladze}
\affil[1]{Javakhishvili State University, 3 Chavchavadze Ave., Tbilisi 0179, Georgia}
\affil[2]{Andronikashvili Institute of Physics, 6 Tamarashvili St., Tbilisi 0177, Georgia}
\affil[3]{Department of Physics, Carnegie Mellon University, PA 15213, USA}

\maketitle

\begin{abstract}
The information conservation principle is probed for classically isolated systems, like the Hubble sphere and black holes, for which the rise of entanglement entropy across their horizons is expected. We accept the analogy of Landauer's principle that entanglement information should introduce some negative potential energy, which corresponds to the positive energy of measurements that destroy this quantum behavior. We estimated these dark-energy-like contributions and found that they can explain the dark energy of the Universe and also are able to resolve the observed superluminal motion and redshift controversies for black holes.

\end{abstract}


\section{Introduction}
\label{Introduction}

The idea that nature is built by information serves as a basis for so-called emergent theories in which classical space-time is not a fundamental concept but can be of information-statistical origin \cite{Car, Ashtekar:1992tm, Connes:1994hv, Jac, Padmanabhan:2009vy, Ver, Cao:2016mst, Amari, Caticha:2010yy}. In general, physical theories are designed for processing information about the world, which suggests that even geometry can be explained in terms of statistical notions \cite{Amari}, i.e. geometry is not an a priori objective reality, but construction of the observer and the metric of space-time represents his potential knowledge (or ignorance) \cite{Caticha:2010yy}. Then an observer's uncertainty grows with the volume of space that comports well with the assumption that the Black Holes (BH) entropy,
\begin{equation} \label{S-BH}
 S_{BH} = \frac{\pi r_s^2}{G} ~,
\end{equation}
is proportional to the area of its Schwarzschild horizon $r_s = 2GM_{BH}$ \cite{Bekenstein:1973ur}. In the emergent spacetime paradigm, (\ref{S-BH}) can be interpreted as  a specific version of the holographic principle -- the equality of the degrees of freedom inside the horizon and on the boundary surface \cite{Padmanabhan:2012ik},
\begin{equation} \label{N=N}
N_{\rm bulk} = N_{\rm surf}~.
\end{equation}
Note that applying this information relation to the Hubble volume with, $N_{\rm bulk} = 2E/k_BT$ and $N_{\rm surf} = 4\pi /G H^2$ (where $E$ is the Komar energy inside the Hubble volume, $k_B$ is Boltzmann's constant and $T = H/2\pi$ is the Gibbon’s Hawking temperature) one could also arrive at the Friedmann equation for a spatially flat universe.

In general, the concept of entropy (or information) is far from being well understood. This is the case of the BH domain, where we are not able to associate the entropy (\ref{S-BH}) with the counting of microstates and for the Unruh effect, where it appears that the number of microstates might be related to the motion of the observer \cite{Wald}. This suggests that a definition of entropy more general than that introduced by Boltzmann, Shannon or von Neumann might exist \cite{Wilde}.

Entropy in classical or quantum theories quantifies two seemingly opposite entities: uncertainty (or ignorance) and information (or order). However, there are deep reasons to expect relations between different kinds of entropies adopting the principle of total entropy conservation \cite{I-cons},
\begin{equation} \label{S=const}
S_{\rm tot} = const~.
\end{equation}
This basic principle of information science, which is generalisation of the relations like (\ref{N=N}), constitutes one of the most important elements to our understanding of the quantum universe. According to (\ref{S=const}), after the erasure of some information about a physical system the thermodynamic component of its environment's entropy increases, while the entropy for the whole system, the information coding device plus its environment, is conserved. To keep the relation (\ref{S=const}) and at the same time obey the second law of thermodynamics, it seems that we need to include some negative components in standard expressions for entropies \cite{Neg-S-1, Neg-S-2, Cerf:1995sa}.

In general, the sign of entropy is conventional. Confusions with the sign can arise in the models concerning complex systems and using some combinations of different types of entropy. An example is von Neumann's measurement scheme, in which the observer uses the “memory” entangled with a quantum system and the joint quantum entropy can be less than the sum of marginal entropies of subsystems. This effect allows the creation of entropy in the measurement device which is counterbalanced by the negative entropy of the quantum system itself \cite{Neg-S-1, Neg-S-2, Cerf:1995sa}. This results in the conservation of entropy in the measurement process and thus can be the key observation in the quantitative realization of the principle of the total entropy conservation (\ref{S=const}).

The concept of entropy of the Universe, $S_U$, also continues to cause confusion within the scientific community \cite{Wallace:2009dc}, because the entropy is best defined for closed systems in thermodynamic equilibrium, but the universe is usually assumed not to have an environment. If the early universe was in equilibrium (as the CMB appears to suggest), already having maximal thermodynamic entropy, it could not have increased later due to the second law of thermodynamics. From the other hand, at the early stages the value of the total 'universal' entropy, $S_U$, seems to be very low. These two facts also suggest that in its classical definition of entropy some negative, non-thermodynamic components are missing. The low value of $S_U$ may be followed from the definition of von Neumann entropy, which is expected to be zero applied for the whole universe since in the quantum mechanical approach the universe is in a (huge) pure state. This is in contradiction with the fact that thermodynamic entropy for the universe is enormous. The significance of these paradoxes has been pointed out long ago \cite{Penrose} and still, there is little consensus about how to define maximal and minimal values and evolution rate of the entropy for the universe \cite{Lin-Ega}. The resolution of these contradictions can be the application of the main principle of the information approach (\ref{S=const}) - the conservation of total entropy.

In our previous papers we have introduced the Information-Probabilistic description of the Universe considering it as the finite ensemble of quantum particles \cite{Gogberashvili:2018jkg, Gogberashvili:2016wsa, Gogberashvili:2013cea, Gogberashvili:2014ora, Gogberashvili:2010em, Gogberashvili:2008jr, Gogberashvili:2007dw}. Here we want to explore the possible consequences of this formalism and thus to probe it. One of the main hypotheses of the approach is based on the applications of the entropy conservation principle (\ref{S=const}) to the physical systems that are quasi-isolated by classical horizons. In General Relativity horizons that block information are unavoidable and for space-time regions, like the Hubble sphere or BHs, different kinds of entropies can be introduced. We claim that entanglement energy terms usually are missed in the equation of states written for such systems.

The paper is organized as follows.  After the introduction, the main ideas of the Information-Probabilistic formalism are presented in Sec.~\ref{I-P}, which suggest the introduction of negative entanglement energy terms for classically isolated systems. As a first example, in Sec.~\ref{Cosmology} we discuss the possible entanglement origin of the dark energy term in cosmological equations for the Hubble sphere. The Sec.~{\ref{BH}} is devoted to BH problems. Based on the information conservation principle, we try to resolve the famous information paradox, apparent superluminal motion, and gravitational redshift variation problems. Finally, in Sec.~\ref{Conclusions} we summarize our results.


\section{Information-Probabilistic idea}
\label{I-P}

The Information-Probabilistic model considers the universe as a superposition of quantum states of a finite ensemble of particles \cite{Gogberashvili:2018jkg, Gogberashvili:2016wsa, Gogberashvili:2013cea, Gogberashvili:2014ora, Gogberashvili:2010em, Gogberashvili:2008jr, Gogberashvili:2007dw}. The total entropy of the universe in this approach can be assumed to be zero since the system is finite and there are absent outside observers. In any quantum measurement process the three systems, quantum object, memory (or apparatus), and the observer are involved. Then in the simplified model, the total entropy of the Universe, $S_U$, can be symbolically written as the sum of information ($I$), thermodynamic ($S$), and quantum (entanglement, ${\cal H}$) components. Then, the assumption (\ref{S=const}) obtains the form \cite{Gogberashvili:2008jr}:
\begin{equation} \label{S=0}
S_U = I + S + {\cal H} = 0~,
\end{equation}
which can be understood as the conservation law for all kinds of information in the Universe.

The expression (\ref{S=0}) seems to be contradictory since the information and entanglement entropies ($I$ and ${\cal H}$) usually are assumed to be dimensionless, while the standard definition of thermodynamic entropy, $S$, contains Boltzmann's constant and is considered to be a dimensional quantity. However, in basic physical laws, Boltzmann's constant always appears together with the absolute temperature (which is related to the mechanical velocities of atoms and molecules) and may be considered as its measure in kelvins \cite{Gogber, Atkins, Ka-Ko}. So, it is natural to relate Boltzmann's constant to the absolute temperature instead, and to consider $S$ in (\ref{S=0}) as a form of Shannon’s dimensionless information entropy that is determined for any system (not necessarily the thermodynamic one) characterized by a certain set of random variables \cite{Inform}. The entanglement entropy, ${\cal H}$, also is a form of information entropy that describes needed information to have full knowledge about other parts of the quantum system.

In the elementary measurement process, the relation (\ref{S=0}) shows the balance of the random $I$ and $S$ systems with the negative entanglement entropy ${\cal H}$. In other words, the mixed state of observer and memory is entangled with the quantum system, which carries negative conditional entropy, such that the combined triple system is always in the pure state. The information interpretation of quantum mechanics only allows unitary time-evolution (including measurement processes), since quantum probabilities arise in unitary time development, thanks to the negative entropy of the unobserved quantum system. In this picture, the expression of the entropy of the whole universe (\ref{S=0}) remains zero at all stages of its evolution.

The condition of entropy neutrality of the universe (\ref{S=0}) suggests information quantization \cite{Zei}, such as the discreteness of the charge might be thought of as the consequence of Gauss' law for the electrically neutral universe -- charge inside of any (even infinitesimally small) volume should be canceled by outside charges. Analogously, there should exist a unit of elementary information (which can be mathematically described by a binary system of numbers) that carries plus (minus) one bit of information.

Since information is physical (according to Landauer’s and Brillouin’s principles), it should be associated with some physical parameter, like energy or mass \cite{Landauer, Info, Ilgin:2014dua}. While there is a well-defined amount of information carried away by the energy flow, the reverse is not necessarily true, some information can be stored in degenerate states and it is possible for the system to lose information without losing energy \cite{Unr}. Also, there is no unique standard of mass that appears additive, e.g. electric charge. Moreover, mass and energy have no polarity and usually are taken to be positive. In \cite{Gogberashvili:2008jr} we had suggested that the convenient physical parameter to measure information (or entropy), instead of mass or energy, is the action that usually is considered to be additive like entropy and have a unique discrete value, $\hbar$. In this approach, the maximal entropy (minimal information) principle \cite{Jay, Pavon:2012qn, B:2017kyi}, is equivalent to the familiar minimal action principle. However, the minimal information principle is more general, especially for quantum descriptions, since in physical equations one can introduce the information entropy terms that correspond to measurement processes.

The relation of the classical action to the thermodynamic entropy translates the condition of entropy neutrality (\ref{S=0}) into the null action principle - the sum of all components of the action for a physical system (including the boundary terms) is zero. So, the action evaluated over the entire (past and future) history of a closed system, in particular for the entire finite universe, should be zero. The motivation came also from the observation that the action constitutes a fundamental physical quantity, respecting all known symmetries and being the basic quantity in deriving the dynamics of most systems. Thus the action of the entire 4-manifold describing the universe should be finite  \cite{Action} and zero \cite{Gogberashvili:2008jr}. One consequence of this principle for an isolated physical system is the zero-energy condition,
\begin{equation} \label{rho-U=0}
\rho_{\rm tot} = 0~,
\end{equation}
i.e. the total density of all the forms of energy associated with this system should be zero at any moment.

To fulfill the condition (\ref{rho-U=0}) for a macroscopic object one needs large negative ingredients, some of which correspond to entanglements of its quantum particles with the outside universe. Entanglements should introduce a certain amount of negative potential energy corresponding to the positive energy needed for the measurements that destroy this quantum behavior. So entanglement energy in a physical vacuum can be estimated using the analog of Landauer's principle:

\vskip 3mm
\noindent
{\it For an observation in which 1 bit of entanglement information is lost the energy that must eventually be emitted into the vacuum is
\begin{equation} \label{Landauer}
E_{\cal H} \geq k_B T_{\rm CMB} \ln 2 \sim 1.6 \times 10^{-13}~GeV~,
\end{equation}
where $k_B \approx 8.6 \times 10^{-5}~ eV/K$ is the Boltzmann constant and $T_{\rm CMB} \approx 2.7~K$ is the present temperature of the vacuum.}
\vskip 3mm

At the end of this section let us note that the consequence of the condition (\ref{rho-U=0}) for the whole finite Universe is that it can emerge without violation of the energy conservation, which appears to be the preferable point of view in cosmology \cite{Fey,Haw-2}.


\section{Entanglement energy for the Hubble sphere}
\label{Cosmology}

Let us first consider the application of the zero-energy condition (\ref{rho-U=0}) of Information-Probabilistic description to cosmology. The Hubble radius (for the spatially flat universe $k = 0$),
\begin{equation} \label{R_H}
R_H = \frac 1H \approx 14.5~Gly~,
\end{equation}
usually is used to fix variations of metric tensors in the action formalism for cosmology. Contrarily, variations of quantum fields should be extended up to the event horizon, which is the real boundary of space-time. As a result, the entanglement energy of quantum particles across the apparent horizon is missed in the cosmological equations written for the Hubble volume. Let us show that, using the null energy assumption for the finite universe (\ref{rho-U=0}), we can identify this missing boundary term in cosmological equations with the dark energy density \cite{Gogberashvili:2018jkg, Lee:2007zq}.

For a homogeneous, isotropic and flat universe there are two independent Friedmann equations with the cosmological term $\Lambda$:
\begin{equation} \label{FRW}
\begin{split}
H^2 &= \frac {8\pi G}{3} \rho + \frac 13 \Lambda~,\\
\dot{H} + H^2 &= - \frac {4\pi G}{3} (\rho + 3 p) + \frac 13 \Lambda~.
\end{split}
\end{equation}
It is known that the system (\ref{FRW}) can be rearranged and written in the form where $\Lambda$ does not appear explicitly:
\begin{equation} \label{FRW-2}
\begin{split}
\dot{H} &= - 4\pi G (\rho + p)~,\\
\dot{\rho} &= - 3 H (\rho + p)~.
\end{split}
\end{equation}
So, if instead of (\ref{FRW}), one will choice (\ref{FRW-2}) as the independent system of cosmological equations, $\Lambda$ obtains the role of an integration constant that can be fixed from an equation of state. Indeed, excluding the combination $(\rho + p)$ from the equations in (\ref{FRW-2}) and integrating over time we find,
\begin{equation} \label{balance}
H^2 = \frac {8\pi G}{3} \rho + C~,
\end{equation}
where the cosmological term re-appears in the form of an arbitrary integration constant $C$, which should be chosen as $C = \Lambda/3$ in order to obtain the first Friedmann equation in (\ref{FRW}).

For the finite universe limited by the event horizon,
\begin{equation} \label{R_e}
R_e = \int_1^\infty \frac {da}{a^2 H(a)} \approx 16.7~Gly~,
\end{equation}
when $H \to 1/R_e$, the energy balance condition (\ref{balance}) should obtain the form:
\begin{equation} \label{}
\frac {1}{R_e^2} = \frac {8\pi G}{3}\rho_U + C ~.
\end{equation}
Using the null-energy condition for closed systems (\ref{rho-U=0}), this equation allows us to fix the integration constant $C$ in (\ref{balance}),
\begin{equation} \label{}
C = \frac {1}{R_e^2}~.
\end{equation}
Then, from the equation (\ref{balance}) written for the Hubble sphere $H \sim 1/R_H$, the estimated value of dark energy density is very close to the observed value,
\begin{equation} \label{DE}
\frac {3C}{8\pi G} = \rho_c \frac {R_H^2}{R_e^2} = 0.75 \rho_c ~,
\end{equation}
where for the critical energy density we use the standard expression
\begin{equation} \label{rho-c}
\rho_c = \frac {3H^2}{8\pi G} \approx 10^{-26}~kg/m^3 \approx 5 \times 10^{-6}~ GeV/cm^3~.
\end{equation}

So, in the Information-Probabilistic approach dark energy and accelerated expansion of the universe are connected with the entanglement of sub-horizon particles with the particles outside of the Hubble sphere.


\section{Entanglement entropy of black holes}
\label{BH}

Another important solution of Einstein's equation, for a region screened by a horizon, represents Black Holes (BHs). It is known that our Universe looks like a BH for an outside observer -- outgoing radial null geodesics cannot escape it, but incoming matter can enter the Hubble sphere. So, we can try to address the Information-Probabilistic description \cite{Gogberashvili:2016wsa} to BHs as well and propose solutions to some related problems.

One famous problem with BH is the information paradox (see, for example \cite{Polchinski:2016hrw}). Within the Information-Probabilistic approach, the total entropy of a BH, according to (\ref{S=0}), can be written as the sum of the informational, thermodynamic and entanglement components. Since a BH classically can be described just by the one known parameter, $M_{BH}$, its information entropy $I$ can be considered to be zero and the entropy balance condition (\ref{S=0}) obtains the simpler symbolic form:
\begin{equation} \label{S+H=0}
S_{BH} + {\cal H}_{BH} = 0~.
\end{equation}
When matter crosses the horizon its classical information decreases for outside observers and the zero-entropy principle (\ref{S+H=0}) dictates that the entanglement entropy ${\cal H}_{BH}$ should increase. This is similar to the case considered in the previous section of the appearance of dark energy in sub-Hubble volume due to the entanglement energy with outside particles. For BHs, this effect would explain the information paradox, since rather than information being lost it would contribute to the entanglement. The exciting part of this observation is that this effect should be stronger near the BH horizon and we can make measurable predictions to check this hypothesis.

Let us estimate the energy corresponding to the BHs entanglement information, ${\cal H}_{BH}$, using the Landauer principle (\ref{Landauer}). The potential energy stored in entanglements should contribute the negative sign into the zero-energy condition (\ref{rho-U=0}) for the case of BH,
\begin{equation} \label{}
\rho_{BH} = 0~.
\end{equation}
We can model this dark-energy-like behavior introducing in the Schwarzschild metric the effective negative cosmological term, ${\cal L}$, for the matter entangled to the BH \cite{Gogberashvili:2018jkg, Lee:2007zq, Gogberashvili:2016xcx}:
\begin{equation} \label{Schwarz}
ds^2 = - \left( 1 - \frac{2GM}{r} + \frac{r^2 {\cal L}}{3} \right)dt^2 + \left( 1 - \frac{2GM}{r} + \frac{r^2 {\cal L}}{3}\right)^{-1} dr^2 + r^2 d \theta^2 + r^2 \sin^2 \theta d \phi^2 ~.
\end{equation}
Of course ${\cal L}$ is not the global cosmological constant for all spacetime, as it was in the previous section (${\cal L} \ne \Lambda$), since it is caused by the entanglement entropy of the BH and the matter.

The expression for velocity of matter particles in the metric (\ref{Schwarz}) has the form \cite{Hackmann:2008zz}:
\begin{equation} \label{v}
\biggl( \frac{dr}{dt} \biggr)^2 = \frac{1}{E^2} \biggl( 1 - \frac{r_s}{r} + \frac{r^2 {\cal L}}{3} \biggr)^2 \biggl[ E^2 - \biggl( 1 - \frac{r_s}{r} + \frac{r^2 {\cal L}}{3} \biggr) \biggl( \epsilon + \frac{L^2}{r^2} \biggr) \biggr]~,
\end{equation}
where $L$ is the angular momentum and the parameter $\epsilon$ is 1 for massive particles and equals 0 for massless ones. For the relativistic jets from BHs, for simplicity in (\ref{v}) we can assume $\epsilon = 0$ and $L = r^2 \phi' = 0$. Then, the expression for the radial velocity of matter (\ref{v}) becomes
\begin{equation} \label{v=}
v_r = 1 - \frac{r_s}{r} + \frac{r^2 {\cal L}}{3} ~,
\end{equation}
which contains the entanglement energy density ${\cal L}$ that imitates the acceleration of space-time. We see from (\ref{v=}) that when
\begin{equation} \label{lambda}
{\cal L} = - \frac{3}{r^2} \biggl( 1 - \frac{r_s}{r} - v_r \biggr) > \frac{3 r_s}{r^3} ~,
\end{equation}
$v_r$ would be greater than one and the distant observer would observe apparent superluminal motion.

To estimate the local value of ${\cal L}$ in (\ref{Schwarz}) and thus probe the Information-Probabilistic description, we want to consider the behavior of matter close to the best-studied supermassive BH in the M87 galaxy, having the mass
\begin{equation} \label{M-BH}
M_{BH (M87)} \approx 6.5 \times 10^9~ M_{\odot} \approx 1.3 \times 10^{40}~kg~,
\end{equation}
and thus, the Schwarzchild radius
\begin{equation} \label{r-s}
r_s \approx 5.9 \times 10^{-4}~pc \approx 1.8 \times 10^{13}~m~.
\end{equation}
It is known that matter close to that BH exhibit apparent superluminal motions for two knots of matter \cite{HST-1}, measured radial velocities and distances from the BH of which are \cite{Snios:2019bqa}:
\begin{equation} \label{v,d}
\begin{split}
v_{\rm HST-1} &\approx 6.3 \pm 0.4~c~, \qquad d_{\rm HST-1} \approx 80~pc \approx 2.5 \times 10^{18}~m~, \\
v_{D} &\approx 2.4 \pm 0.6~c, \qquad d_{D} \approx 226~pc \approx 7.0 \times 10^{18}~m~.
\end{split}
\end{equation}
Using (\ref{r-s}) and the data (\ref{v,d}) on locations of knot HST-1 and knot D in 2017, from (\ref{lambda}) we can estimate the required values of effective ${\cal L} \sim 3 r_s/d^3$ that could explain the above mentioned behaviour of knots,
\begin{equation} \label{L}
{\cal L} \sim 10^{-41}~ m^{-2}~.
\end{equation}
Let us estimate entanglement entropy and its distribution around the BH of M87 and compare the associated effective cosmological constant term with (\ref{L}).

We can find the value of classical entropy of the BH in M87 using (\ref{S-BH}),
\begin{equation} \label{S=}
S_{BH} \approx \frac{\pi r_s^2}{l_{p}^2} \sim 10^{96}~,
\end{equation}
where $r_s$ is done in (\ref{r-s}) and $l_{p} \approx 1.6 \times 10^{-35}~m$ denotes Planck's length. Then from (\ref{S+H=0}) for the entanglement entropy of the BH we have
\begin{equation} \label{H}
    {\cal H}_{BH} \sim - 10^{96}~.
\end{equation}
At the moment of creation of the BH, most of the particles from the surrounding region are expected to entangle with the inside matter. Later they would spread out, mostly due to the expansion of the Universe and we need to evaluate the radius of the sphere, where those particles are spread. We need also to estimate the total number of particles (and their energy $E_{\rm tot}$), to which the entanglement entropy (\ref{H}) was shared outside the BH and compare it with the number of entangled particles in an object of mass $M$ close to the BH horizon. Then the portion of the total entanglement entropy (\ref{H}) for this object will be
\begin{equation} \label{H=-H}
    {\cal H} = - {\cal H}_{BH} \frac{M}{E_{\rm tot}}~.
\end{equation}

The exact age of the BH in M87 is currently unknown, but the age of the home galaxy is about 13.2 billion years. To calculate the current physical radius of this BH creation zone, $R$, we use the scale factor
\begin{equation}
a = \frac{13.2}{13.8} \approx 0.96~,
\end{equation}
which gives,
\begin{equation}
    R = \int_{0}^{0.96} \frac{da}{a} \frac{1}{aH(a)} \sim 10^4 ~ Mpc~.
\end{equation}
The total energy within this large radius can be roughly evaluated using the value for the critical density of the universe (\ref{rho-c}),
\begin{equation} \label{E-tot}
    E_{\rm tot} \approx \frac 43 \pi R^3 \rho _{c} \sim 10^{54}~kg~.
\end{equation}

Let us estimate the entanglement energy for one of the knots in M87, for example for Knot HST-1 in (\ref{v,d}). We can write the expression of its total mass as
\begin{equation} \label{M-HST}
M_{\rm HST-1} = \rho_{BH} V_{\rm HST-1}~,
\end{equation}
where $\rho_{BH}$ denotes the observed central energy density of M87 close to its supermassive BH, which has at least the order \cite{EHT},
\begin{equation} \label{rho}
\rho_{BH} \sim 10^{-17} ~ kg/m^3~.
\end{equation}
Inserting (\ref{H}), (\ref{E-tot}) and (\ref{M-HST}) into (\ref{H=-H}) we find the value of the entanglement entropy for the Knot HST-1,
\begin{equation} \label{H-HST}
   {\cal H}_{\rm HST-1} \sim - 10^{25}~ V_{\rm HST-1}~ m^{-3}~,
\end{equation}
and for its total entanglement energy it follows:
\begin{equation} \label{E-HST}
   E_{\rm HST-1} \sim 10^{11}~V_{\rm HST-1}~GeV/m^{3} ~.
\end{equation}
Finally, using for the expression for the Newton constant $G = l_p/m_p$ ($m_p \approx 10^{19}~GeV$ is Planck's constant), for the dark-energy-like term corresponding to Knot HST-1 we obtain,
\begin{equation} \label{L-HST}
   {\cal L}_{\rm HST-1} = 8\pi G  \rho_{\rm vac} \approx \frac {8\pi l_p} {m_p } \frac{ E_{\rm HST-1}}{V_{\rm HST-1}} \sim 10^{-41}~m^{-2}  ~,
\end{equation}
which is of order of the required values (\ref{L}) that explains apparent superluminal motions of knots close to the BH in M87.

At the end of this section let us consider also the quasars redshifts problem \cite{Arp}. According to (\ref{S-BH}), a solar mass BH has an entropy $\sim 10^{77}$ that is 20 orders of magnitude larger than the thermodynamic entropy of the sun. This additionally emphasizes our motivation, since it clearly shows that one should not think of BH entropy as just the entropy of matter that fell into it. The faster a BH grows, the bigger is a discrepancy between its total and thermodynamic entropies, the contribution to the entanglement entropy is more significant and can result in observable effects. The fastest-growing BH yet known, $\sim 1\%$ growth rate every million years, is the quasar QSO SMSS~J215728.21-360215.1 of mass $\sim 10^{10}~M_\odot$ \cite{Wolf:2018ivm}. Thus, thermodynamic and entanglement entropies of that BH would increase by $10^8 \times 10^{77}$ per million years. Variations of entanglement entropy close to a BH horizon will contribute to the dark-energy-like term that modifies $g_{tt}$ in (\ref{Schwarz}) and leads to the additional redshift of the outgoing radiation.

According to accepted interpretations, quasars are believed to be the fastest and farthest objects. However, several observations indicate that this velocity and distance interpretation is flawed \cite{QSO}. We note that it is not correct to interpret the redshift of quasars in terms of cosmological redshift alone, the contribution of the gravitational redshift must be considered too. Gravitational redshift usually is ignored, since it is estimated to be about two orders of magnitude smaller than Doppler shift due to random motion. In our model, due to the entanglement energy, the $g_{tt}$ component in (\ref{Schwarz}) is modified close to the horizon of the BH and the formula for gravitational redshift  $\lambda ' = \lambda \sqrt{g_{tt}}$ will dictate modification of redshifted wavelength $\lambda'$,
\begin{equation} \label{z}
\frac{\lambda'}{\lambda} = \sqrt{ 1 - \frac{2GM}{r} + \frac{r^2 {\cal L}}{3}}~,
\end{equation}
since now we have an additional term $\frac{r^2 {\cal L}}{3}$. One can notice that, even for the relatively small value in (\ref{L-HST}), the entanglement energy term with ${\cal L}$ in the definition of the redshift (\ref{z}) becomes dominant for the $r \sim 1~kpc$ region close to a supermassive BH.


\section{Conclusions}
\label{Conclusions}

To summarize, we have suggested a way to probe the information conservation proposal (\ref{S=0}) in cosmology and for black holes. According to this description, the information which seemed to be lost for external observers should be manifested as the entanglement entropy, which reproduces the dark-energy-like terms and thus additional energy near their horizons. In the case of the Hubble horizon, this idea can explain the observed dark energy. For black holes, dark-energy-like terms neatly explain apparent superluminal motion, reproducing the data for the well-known supermassive black hole in M87 galaxy and provide qualitative explanation for the anomalous redshifts of far quasars.


\section*{Statements and Declarations}
\subsection*{Funding}

M. G. acknowledges the support by Shota Rustaveli National Science Foundation of Georgia through the grant DI-18-335.

\subsection*{Competing Interests}

The authors have no relevant financial or non-financial interests to disclose.

\subsection*{Author Contributions}

Both authors contributed to the research and  read and approved the final manuscript.


\end{document}